\documentclass[aps,prd,twocolumn,showpacs,floatfix,superscriptaddress]{revtex4-2}
\usepackage{graphicx}
\usepackage{dsfont}
\usepackage{xcolor}
\usepackage{multirow}
\usepackage[normalem]{ulem}
\usepackage{enumitem} 
\usepackage{lineno}
\usepackage{bm}
\usepackage{slashed}
\usepackage{footnote}
\definecolor{refcol}{RGB}{34,34,178}%{178,34,34}%{0,100,170}%{0,0,205}
\usepackage[colorlinks,linkcolor=blue,citecolor=blue, urlcolor=blue]{hyperref}
\usepackage{microtype}
\usepackage{float}
\usepackage[caption = false]{subfig}
\usepackage{appendix}
\usepackage{soul}
\usepackage{amsmath}
\usepackage{bbold}
\graphicspath{{./Fig/}}

\newcommand{\nn}{\nonumber}
\newcommand{\RNum}[1]{\uppercase\expandafter{\romannumeral #1\relax}}

\begin{document}

\title{Vector interaction bounds in NJL-like models from LQCD estimated curvature of the chiral crossover line}

\date{ \today }
\author{Mahammad Sabir Ali}
\email{sabir@niser.ac.in}
\affiliation{School of Physical Sciences, National Institute of Science Education and Research, An OCC of Homi Bhabha National Institute, Jatni-752050, India}

\author{Deeptak Biswas}
\email{deeptakb@niser.ac.in}
\affiliation{School of Physical Sciences, National Institute of Science Education and Research, An OCC of Homi Bhabha National Institute, Jatni-752050, India}

\author{Chowdhury Aminul Islam}
\email{chowdhury@physik.uni-frankfurt.de}
\affiliation{Institut f\"{u}r Theoretische Physik, Johann Wolfgang Goethe–Universit\"{a}t, Max-von-Laue-Str. 1, D–60438 Frankfurt am Main, Germany}
\affiliation{Center for Astrophysics and Cosmology, University of Nova Gorica, Vipavska 13, SI-5000
Nova Gorica, Slovenia}
 
%%%%%%%%%%%%%%%%%%%%%%%%%%%%%%%%%%%%%%%%%%%%%%%%%%%%%%%%%%%%%%%%%%%%%%%%
\begin{abstract}
We obtain improved bounds on both the flavor-independent and -dependent vector interactions in a $2+1$-flavor Nambu\textendash Jona-Lasinio (NJL) model using the latest precise LQCD results of the curvature coefficients of the chiral crossover line. We find that these lattice estimated curvature coefficients allow for both attractive and repulsive types of interactions in both the cases. With this constrained ranges of vector interactions, we further predict the behavior of the second $(\kappa_2^B)$ and fourth  $(\kappa_4^B)$ order curvature coefficients as a function of the strangeness chemical potential $(\mu_S)$. We observe that the flavor mixing effects, arising from the flavor-independent vector interaction as well as from the 't Hooft interaction, play an important role in $k_2^B$. We propose that the mixing effects due to the vector interaction can be separated from those arising from the 't Hooft interaction by analyzing the behavior of $k_2^B$ as a function of $\mu_S$. Finally, we locate the critical endpoint in the $T-\mu_B$ plane using the model-estimated ranges of vector interactions and find the model's predictions to be consistent with the latest LQCD bounds. 
\end{abstract}
%%%%%%%%%%%%%%%%%%%%%%%%%%%%%%%%%%%%%%%%%%%%%%%%%%%%%%%%%%%%%%%%%%%%%%%%
%For fun, see the PACS list at  https://www.aip.org/publishing/pacs/pacs-2010-regular-edition
%\pacs{47.15.-x}
%%%%%%%%%%%%%%%%%%%%%%%%%%%%%%%%%%%%%%%%%%%%%%%%%%%%%%%%%%%%%%%%%%%%%%%%%%%%%%%%%%%%%%%%%%%%%%%%%%%
\maketitle
\pagenumbering{arabic} 
%%%%%%%%%%%%%%%%%%%%%%%%%%%%%%%%%%%%%%%%%%%%%%%%%%%%%%%%%%%%%%%%%%%%%%%%
\section{Introduction}
%%%%%%%%%%%%%%%%%%%%%%%%%%%%%%%%%%%%%%%%%%%%%%%%%%%%%%%%%%%%%%%%%%%%%%%%
Over the last few decades, substantial progress has been made in understanding the strong interaction system at low energies, where the spontaneous breaking of (approximate) chiral symmetry plays a crucial role. This broken symmetry is restored as the temperature $(T)$ and/or baryon chemical potential $(\mu_B)$ of the system increases. The transition from the broken to the restored symmetry phase is crucial for understanding phenomena such as the quark-gluon plasma (QGP) produced in relativistic heavy ion collisions (HICs), as well as the early universe in the context of the Big Bang and compact astrophysical objects like neutron stars, which may have a quark core~\cite{Annala:2019puf}.

First-principle lattice QCD (LQCD) calculations have made significant progress over the past two decades, greatly enhancing our understanding of the chiral transition at zero or low baryon chemical potential $(\mu_B)$. However, the lattice artifacts arising from the well-known \textit{sign problem} restrict the extension of such studies at large $\mu_B$. In the current scenario, techniques such as Taylor expansion around $\mu_B=0$~\cite{Gavai:2003mf, Gavai:2004sd, HotQCD:2018pds} or analytical continuation from imaginary $\mu_B$~\cite{Bonati:2015bha, Borsanyi:2020fev} have produced reliable results for $\mu_B/T \leq 3$~\cite{Borsanyi:2021hbk,Philipsen:2021qji}. Additionally, functional methods like the Dyson-Schwinger equations (DSE)~\cite{Fischer:2018sdj} and functional renormalization group (FRG)~\cite{Dupuis:2020fhh} offer trustworthy results for higher $\mu_B$ values through their improved truncation schemes. 

On the other hand, effective models constructed based on the underlying symmetries of the theory exhibit broad applicability across the $T-\mu_B$ plane. In scenarios where reliable first-principle calculations are lacking, these models can provide significant insights, specifically into the low-energy sector of QCD. Even with advancements in first-principle results, these models continue to be relevant, particularly for exploring extreme conditions such as those characterized by high baryon density \cite{Buballa:2003qv}.

The Nambu\textendash Jona-Lasinio (NJL) model serves as a quintessential example of an effective model, extensively used to investigate the breaking and restoration of chiral symmetry, as well as the influence of external parameters on the underlying dynamics~\cite{Klevansky:1992qe, Miransky:2015ava}. Initially proposed by Y. Nambu and G. Jona-Lasinio with nucleon degrees of freedom~\cite{Nambu:1961tp, Nambu:1961fr}, the model was later formulated in terms of quark degrees of freedom following the discovery of quarks~\cite{Hatsuda:1994pi, Rehberg:1995kh}. As a low-energy model, one can consider all possible interactions up to a given order, constrained by the underlying symmetries. Even with all the possible terms within the mean-field approximation, the low-energy dynamics can still be captured using only a few parameters, as demonstrated in Ref.~\cite{Gupta:2017gbs}.

Among the various types of interactions, vector interaction plays a crucial role. Its strength can be generally fixed from the vector mesons as they emerge as peaks in the quark–antiquark spectral functions~\cite{Vogl:1991qt,Klevansky:1992qe}. But the masses of these vector mesons are less robustly constrained than those of pseudoscalar mesons in an effective model scenario, since they typically lie close to the model cutoff~\cite{Klimt:1989pm}. Along with this, the vector meson modes in the NJL model are embedded in an unphysical quark--antiquark continuum arising from the lack of confinement. Consequently, the lowest-lying vector mesons, such as the $\rho$, can decay into this continuum. As a result, the model does not reliably reproduce the vector meson mass spectrum as it does for the pseudoscalar sector~\cite{Vogl:1991qt}. Although in these studies, the extracted vector coupling had a value $G_{V}\sim (1.5)\,G_{S}$ depending on the chosen parameter set. This provides one of the earliest quantitative determinations of the interaction strength~\cite{Klimt:1989pm, Vogl:1991qt}. 

Most importantly, in a dense medium, it becomes essential to account for a non-zero vector interaction due to its direct coupling with the number density operator, $(\bar\psi\gamma_0\psi)$. As a result, even if we start with a zero strength of the vector interaction, the non-zero density environment induces a finite strength for this interaction. As it becomes relevant only at finite density, this induced vector interaction cannot be determined using vector meson properties in the vacuum, and its strength remains an uncertain quantity~\cite{Fukushima:2008wg,Abuki:2009ba}. Ref.~\cite{Chu:2014pja} introduces a separate isovector-vector interaction to differentiate between the isoscalar-vector $(\omega)$ and isovector-vector $(\rho)$ channels. However, the strength of these two interactions are treated as a free parameter and varied as a ratio to the scalar coupling constant. Consequently, in much of the existing literature, the strength of the vector interaction ($G_V$) is varied in units of the scalar interaction strength $(G_S)$~\cite{Bratovic:2012qs, Contrera:2012wj, Friesen:2014mha, Contrera:2016rqj,Steinheimer:2014kka,Islam:2014sea,Islam:2015koa}. It is also worth noting that the sign of the induced vector interaction is not universally agreed upon and can be either positive or negative, leading to repulsive or attractive interactions, respectively, as elaborated well in Ref.~\cite{Fukushima:2008wg}. 

The vector interaction has significant effects on the restoration of chiral symmetry at non-zero baryon chemical potential ($\mu_B$) and directly influences the location of the critical endpoint (CEP) in the $T-\mu_B$ plane. Furthermore, the curvature coefficients of the chiral crossover line around zero $\mu_B$ are strongly affected by the vector interaction. These features, particularly the curvature coefficients, have been exploited to understand the role of vector interaction in QCD and to constrain its range in effective model frameworks, and they form a major part of the present study. Refs.~\cite{Kashiwa:2011td,Bratovic:2012qs, Contrera:2012wj, Friesen:2014mha, Steinheimer:2014kka, Contrera:2016rqj} used different effective model frameworks, including variants of the NJL model, to estimate the vector interaction strength primarily from the LQCD-derived curvature of the chiral crossover line, among other observables. Due to the lack of precise estimates from LQCD, the bounds from the curvatue coefficients remain broad but predominantly suggest a repulsive interaction, with the chiral transition becoming a crossover throughout the phase diagram at a sufficiently strong positive interaction strength~\cite{Fukushima:2008wg}. Vector interactions also have a profound impact on bulk thermodynamic properties at high densities, contributing to an acceptable equation of state (EoS) for astrophysical objects, such as neutron stars~\cite{Weissenborn:2011ut} or hybrid stars~\cite{Shao:2013toa,Klahn:2013kga}.

In this article, we employ a $2+1$-flavor NJL model to revisit the problem of determining the strength of the vector interaction, utilizing the improved and more precise LQCD data on the curvature~\cite{HotQCD:2018pds}. Our motivation is drawn from a recent study~\cite{Ali:2024nrz}, which found that the curvature coefficients of the crossover line obtained within a $2+1$-flavor NJL model align remarkably well with LQCD simulations and exhibit minimal dependence on model parameters. Additionally, it was proposed that determining the second-order curvature coefficient of the $T-\mu_B$ line ($\kappa_2^B$) as a function of the strangeness chemical potential ($\mu_S$) on the lattice would enhance the predictability of NJL-type models. We extend this framework by introducing both flavor-independent and flavor-dependent vector interactions, with coupling strengths $G_V$ and $g_V$, respectively.

%~~~
We varied the vector interaction strengths $G_V$ and $g_V$ in units of $G_S$ across positive and negative ranges and evaluated the corresponding $\kappa_2^B$ and $\kappa_4^B$. Using precise lattice QCD data for $\kappa_2^B$, we narrowed down the range of these vector interaction strengths. However, $\kappa_4^B$ estimates from LQCD are consistently zero within uncertainties, offering limited guidance. Lattice results that impose the strangeness neutrality condition (zero net strange quark) provide an additional method for constraining the data with a finite strangeness chemical potential, $\mu_S$. This motivates further exploration of the impact of $\mu_S$ on the curvature coefficients $\kappa_{2,4}^B$ in the presence of vector interactions, previously studied in Ref.~\cite{Ali:2024nrz} without them. This analysis enhances our understanding of flavor mixing between the light and strange quark sectors. In the NJL model, the 't Hooft determinant term generally captures flavor mixing, while a flavor-independent vector interaction couples the two sectors via their number densities as well. By examining the variation of $\kappa_2^B$ with $\mu_S$, we propose a novel approach to distinguish the contributions from these two mechanisms of flavor mixing and outline potential implications. With the narrowed range of $G_V$ and $g_V$, we finally determine the location of the critical endpoint (CEP) in our model, which is consistent with the existing LQCD bound~\cite{Borsanyi:2021hbk,Philipsen:2021qji}, but differs from the current predictions of functional methods~\cite{Gunkel:2021oya,Gao:2020fbl}.

The article is presented in the following way. In section~\ref{sec:form}, we provide the details of the NJL model as required for the present study and briefly discuss the procedure to estimate the curvature. In section~\ref{sec:results}, we show the results and discuss the estimation of the curvature coefficients in different physical conditions. Finally, in section~\ref{ssec:res_cep}, we find the locations of the CEP within the allowed range of parameters before we conclude in section~\ref{sec:sum}.

%%%%%%%%%%%%%%%%%%%%%%%%%%%%%%%%%%%%%%%%%%%%%%%%%%%%%%%%%%%%%%%%%%%%%%%%
\section{Formalism}
\label{sec:form}
%%%%%%%%%%%%%%%%%%%%%%%%%%%%%%%%%%%%%%%%%%%%%%%%%%%%%%%%%%%%%%%%%%%%%%%%
\subsection{Details of the model}
\label{ssec:mod}
We start with the Lagrangian density of a $2+1$-flavor NJL model, including  the vector interaction, which can be written as~\cite{Masuda:2012ed, Fukushima:2008wg, Grunfeld:2024ihq}
\begin{align}
\mathcal{L} & = 
\bar{q}\left(i \slashed{\partial}-\hat{m}+\gamma^{0}\hat{\mu}\right) q  
+ G_S \sum_{a=0}^{8} \left[ (\bar{q} \lambda_a q)^2 
+ (\bar{q} i \gamma^5 \lambda_a q)^2 \right] \nn\\
&- 8 K \left[ \det( \bar{q} P_R q ) 
+ \det( \bar{q} P_L q ) \right] \nn\\
&- \left\{ 
\begin{array}{ll}
G_V(\bar q \gamma^\mu q)^2 \\[1ex]
g_V \sum_{a=0}^8 \left[
(\bar{q} \gamma^\mu \lambda_a q)^2 +
(\bar{q} i \gamma^\mu \gamma^5 \lambda_a q)^2
\right]
\end{array}
\right.
\label{eq:NJL_Lagrangian}
\end{align}
where the quarks are represented by $q$ as $q^{\text{T}}=(u,d,s)$. $\hat{m}=\text{diag}(m_{u},m_{d},m_{s})$ and $\hat{\mu}=\text{diag}(\mu_{u},\mu_{d},\mu_{s})$ are the mass and chemical potential matrices in the flavor space, respectively. $G_S$ is the strength of the Lorentz scalar interaction that is invariant under global $U(3)\times U(3)$ symmetry, where $\lambda_{a}$'s are the Gell-Mann matrices, the generator of $U(3)$ symmetry in the fundamental representation with normalization $\text{Tr}[\lambda_{a}\lambda_{b}]=2\delta_{ab}$ and $\lambda_{0}\propto\mathbb{1}_{3\times3}$. To take into account the effect of axial anomaly, one explicitly breaks the $U(1)_{A}$ symmetry by introducing the Kobayashi–Maskawa–'t Hooft (KMT) interaction~\cite{Kobayashi:1970ji,tHooft:1986ooh} with coupling $K$, where $P_{L/R}=(1\mp \gamma_{5})/2$ is the chiral projection operator. We have further considered two types of vector interaction\textemdash\, flavor independent and flavor dependent~\cite{Masuda:2012ed}. In that regard, they are two different models, and we call them Model-\RNum{1} (flavor independent, $G_V$) and Model-\RNum{2} (flavor dependent, $g_V$). Next, we utilize the meanfield approximation to substitute the auxiliary bosonic fields, obtained through the Hubbard-Stratonovich transformation of the Lagrangian in Eq.~\eqref{eq:NJL_Lagrangian}, with the corresponding expectation values.

From the meanfield Lagrangian, one can obtain the grand canonical potential evaluating the partition function. To introduce temperature, we have used the Matsubara formalism~\cite{Matsubara:1955ws,Kapusta:2006pm,Haque:2024gva}, which connects temperature to the zeroth component of the four-momentum. The free energy can be written as
\begin{equation}
\Omega (T, \mu) = \Omega_{\text{cond}} + \Omega_{\text{vac}} + \Omega_{\text{med}},
\label{eq:pot}
\end{equation}
where $\Omega_{\text{cond}}$ represents the condensation energy, $\Omega_{\text{vac}}$, the vacuum energy or zero-point energy and $\Omega_{\text{med}}$ denotes the medium contribution. The explicit expressions are
\begin{align}
\Omega_{\text{cond}} & = 2G_S \sum_{i} \sigma_i^2 - 4 K \prod_{i} \sigma_i -\begin{cases}G_{V}\left(\sum_{i}n_{i}\right)^{2}\\
g_{V} \sum_{i}n_{i}^{2}
\end{cases}, \nn \\
\Omega_{\text{vac}} &= -2N_c\sum_{i}\int_{\Lambda}\frac{d^3p}{(2\pi)^3}E_i(p)\;\;\; \mathrm{and} \nn \\
\Omega_{\text{med}} &= 
- 2N_c T\sum_{i} \int_0^\infty \frac{ d^3p }{(2\pi)^3}\left( \ln \left[ 1 + e^{-(E_i(p)+\tilde\mu_i)/T} \right]\right.\nn\\ 
&\left. + \ln \left[ 1 + e^{-(E_i(p)-\tilde\mu_i)/T} \right]\right),
\label{eq:pot_indv}
\end{align}
with $N_c=3$, the number of colors; $E_{i}=\sqrt{p^{2}+M_{i}^{2}}$, $\sigma_{i}=\langle\bar{q}_{i} q_{i}\rangle$ and $n_{i}=\langle{q}_{i}^{\dagger} q_{i}\rangle$ are the energy, quark condensate and number density of the $i$-th quark, respectively. To render finite contribution from the vacuum term ($\Omega_{\text{vac}}$), we use $3$-momentum cutoff, $\Lambda$, which also represents the scale of the theory. For $i\ne j\ne k\in \{u,d,s\}$, the $i-$th quark effective mass, $M_{i}$, is  given by the gap equation
\begin{equation}
M_{i} = m_{i} - 4G_S \sigma_{i} + 2K \sigma_{j} \sigma_{k},
\label{eq:mass}
\end{equation}
and the effective chemical potentials are given by
\begin{equation}
    \tilde\mu_{i}=\begin{cases}\mu_{i}-2G_{V}\sum_{j}n_{j}\\
                \mu_{i}-2g_{V} n_{i}\,.
                \end{cases}.
    \label{eq:effective_mu}
\end{equation}
One should immediately note the mixing among different flavors through the effective chemical potential in Model-\RNum{1}, which is absent in Model-\RNum{2}. Consequently, in Model-\RNum{2}, the quark chemical potential $(\tilde\mu_i)$ is modified only by the respective number density $(n_i)$.

To solve for the mean fields, we need to minimize the free energy by solving the gap equations simultaneously
\begin{eqnarray}
    \frac{\partial\Omega}{\partial\sigma_u}=\frac{\partial\Omega}{\partial\sigma_d}=\frac{\partial\Omega}{\partial\sigma_s}=0.
    \label{eq:gap_eq}
\end{eqnarray}
In the present study, we have considered an exact isospin symmetry (ignoring the electric charges of individual flavors), which implies that $\sigma_{u}=\sigma_{d}=\sigma_l$. Hence, the individual quark chemical potentials can be expressed in terms of the baryon $(\mu_B)$ and strangeness $(\mu_S)$ chemical potentials as
\begin{equation}
\begin{split}
    \mu_{u}&=\mu_{d}=\frac{1}{3}\,\mu_{B}\; \mathrm{and} \\
    \mu_{s}&=\frac{1}{3}\,\mu_{B}-\mu_{S} \,,
\end{split}
\label{eq:chemicalpotentials}
\end{equation}
where, $\mu_u$, $\mu_d$ and $\mu_s$ are chemical potentials for $u$, $d$ and $s$ quarks, respectively. 

%----------------------------------------------------------------------------------
\begin{table}[h]
    \centering
    \begin{tabular}{|c|c|c|c|c|c|}
        \hline
        & ~$m_{l}$(MeV)~ & ~$m_s$(MeV)~ & ~$\Lambda$ (MeV)~ & ~$G_S \Lambda^2$~ & ~$K \Lambda^5$~ \\
        \hline
         Set I & 5.5 & 135.7 & 631.4 & 1.835 & 9.29  \\
        \hline
         Set II & 5.5 & 140.7 & 602.3 & 1.835 & 12.36  \\
        \hline
    \end{tabular}
    \caption{Parametrization used in the present work are from Ref.~\cite{Hatsuda:1994pi} (Set \RNum{1}) and Ref.~\cite{Rehberg:1995kh} (Set \RNum{2}).}
    \label{tab:parameter}
\end{table}
%-----------------------------------------------------------------------------------

\subsection{Curvature calculation}
\label{ssec:cur_cal}
The crossover line for the $T-\mu_X$ plane can be expressed with the following ansatz for smaller values of $\mu_X$ ~\cite{Bellwied:2015rza, Bonati:2018nut, HotQCD:2018pds}:
%------------------------------------------------------------------------
\begin{equation}
\label{eq:curvature}
 \frac{T_{pc}(\mu_X)}{T_{pc}(0)}=1-\kappa_{2}^{X} \left(\frac{\mu_X}{T_{pc} (0)}\right)^2 - 
 \kappa_{4}^{X} \left(\frac{\mu_X}{T_{pc} (0)}\right)^4~.
\end{equation}
%------------------------------------------------------------------------
Here, $X$ denotes conserved charges like baryon charge $B$, electric charge $Q$, and strangeness $S$. 
The curvature coefficients $\kappa_{2}^B$ and $\kappa_{4}^B$ in the $T-\mu_B$ plane have been explored in lattice QCD, with Taylor expansion method~\cite{HotQCD:2018pds} and by analytically continuing imaginary chemical potential results to the real axis~\cite{Bellwied:2015rza, Bonati:2015bha, Borsanyi:2020fev}. There is very good agreement between the results obtained using these two techniques. There have also been results for the same within other model frameworks such as the NJL model~\cite{Gupta:2017gbs}, PNJL model~\cite{Bratovic:2012qs, Contrera:2012wj, Friesen:2014mha, Contrera:2016rqj, Steinheimer:2014kka}, perturbative QCD~\cite{Haque:2020eyj}, hadron resonance gas (HRG) model~\cite{Biswas:2022vat, Biswas:2024xxh} and quark-meson model~\cite{Fu:2019hdw, Schaefer:2004en, Braun:2011iz, Fischer:2012vc, Pawlowski:2014zaa, Fischer:2014ata}.

%%%%%%%%%%%%%%%%%%%%%%%%%%%%%%%%%%%%%%%%%%%%%%%%%%%%%%%%%%%%%%%%%%%%%%%%%
\section{Results}
\label{sec:results}
%%%%%%%%%%%%%%%%%%%%%%%%%%%%%%%%%%%%%%%%%%%%%%%%%%%%%%%%%%%%%%%%%%%%%%%%%

\begin{figure}[h!]
\begin{center}
  \includegraphics[scale=0.56]{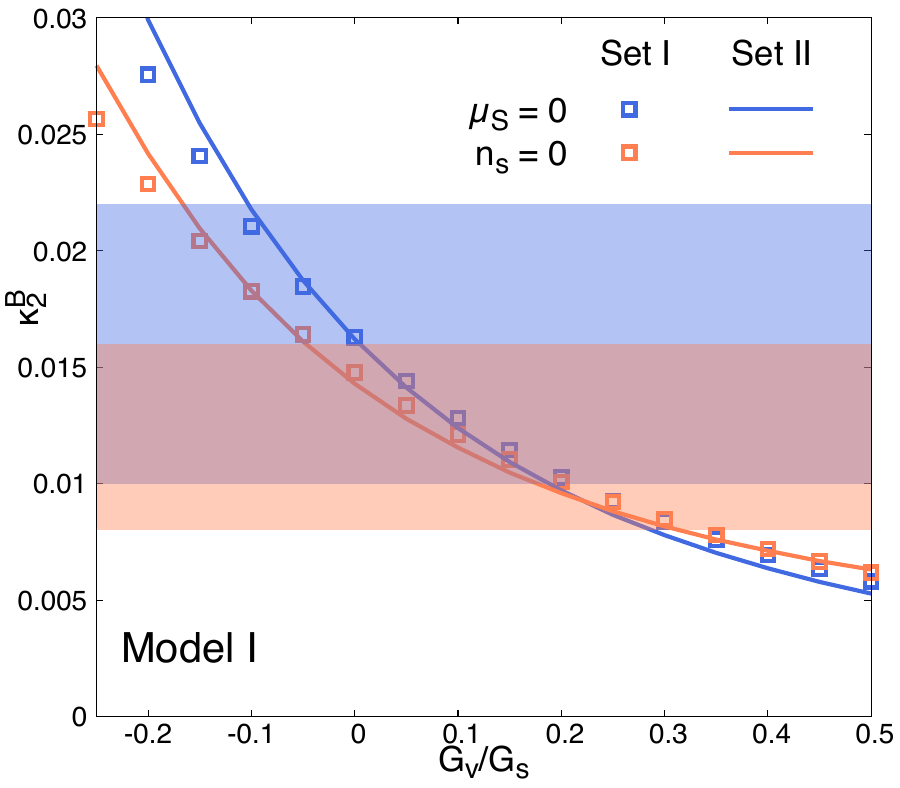}
  \includegraphics[scale=0.56]{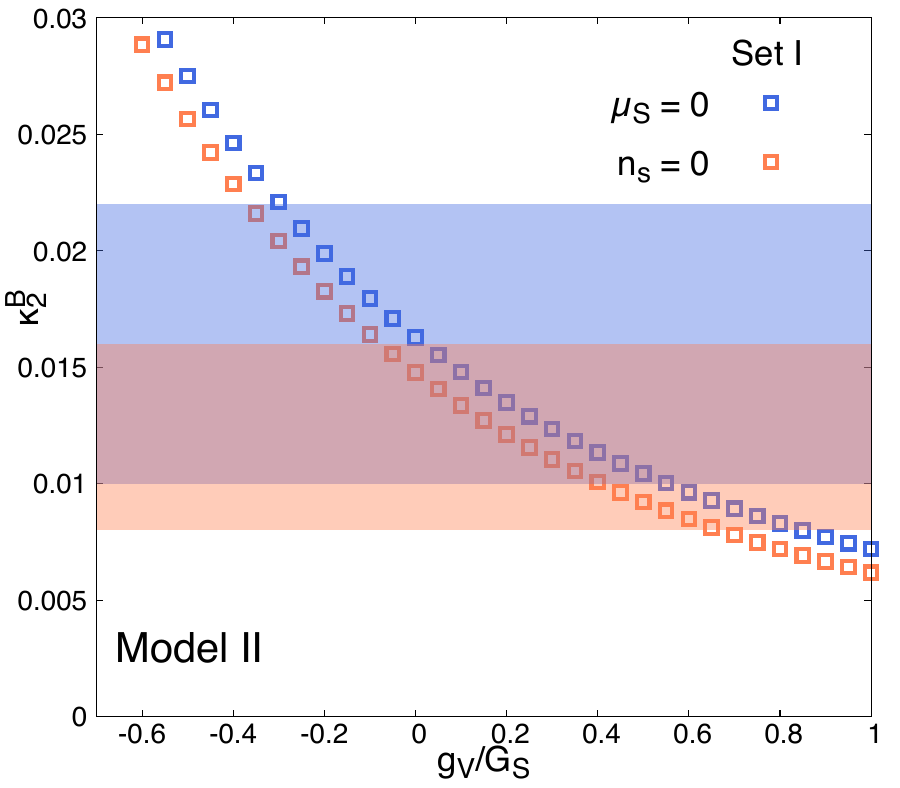}
  \caption{The curvature coefficient ($\kappa_2^B$) as a function of the strengths of vector interactions. Open squares (lines) are obtained using Parameter Set \RNum{1} (Set \RNum{2}), respectively. The bands represent the corresponding Lattice QCD estimations from Ref.~\cite{HotQCD:2018pds}.}
  \label{fig:kappa2_v}
\end{center}  
\end{figure}
%----------------------------------------------------------------------------
%%%%%%%%%%%%%%%%%%%%%%%%%%%%%%%%%%%%%%%%%%%%%%%%%%%%%%%%%%%%%%%%%%%%%%%%%%%%%
\subsection{Constraining $G_V$ and $g_V$ from $\kappa_2^{B}$}
\label{ssec:kappa_vs_vec}
%%%%%%%%%%%%%%%%%%%%%%%%%%%%%%%%%%%%%%%%%%%%%%%%%%%%%%%%%%%%%%%%%%%%%%%%%%%%%
Let us first consider the curvature coefficients of the crossover line evaluated with the inclusion of vector interactions. This analysis aids in constraining the interaction strengths $G_V$ and $g_V$ by comparing them with the lattice QCD estimates of $\kappa_2^{B}$ and $\kappa_4^{B}$, which are available for the crossover line at small values of $\mu_B$. In the NJL model, the pseudocritical temperature ($T_{pc}$) for a given set of $\mu_B$ and $\mu_S$ values is determined as the inflection point of the order parameter, the light quark condensate, $\sigma_l$\footnote{It is to be noted that Ref.~\cite{HotQCD:2018pds} uses a combination of light and strange quark condensates to define the order parameter. This is done to get rid of the divergences that arise in lattice QCD calculation~\cite{Cheng:2007jq}.}. We extract the curvature coefficients $\kappa_2^B$ and $\kappa_4^B$ by parameterizing the respective crossover lines using the ansatz given in Eq.~\eqref{eq:curvature}, within the range $\mu_{B}/T_{pc}(0) \leq 1.0$.

%------------------------------------------------------------------------
\begin{table}[h]
    \centering
    \begin{tabular}{|c|c|c|c|}
        \hline
         & ~$\kappa_2^{B,\mu_S = 0}$ ~ & ~$\kappa_2^{S,\mu_B = 0}$~ & $\kappa_2^{B,n_s=0}$ \\
        \hline
        NJL, Set \RNum{1} & 0.0163 & 0.0134 & 0.0148 \\
        \hline
        NJL, Set \RNum{2} & 0.0162 & 0.0172  & 0.0143 \footnote{The numerical value mentioned in Ref.\cite{Ali:2024nrz} is different as that was calculated with less accuracy.} \\
        \hline
        Lattice QCD~\cite{HotQCD:2018pds} & 0.016(6) & 0.017(5) & 0.012(4)  \\
        \hline
    \end{tabular}
    \caption{Estimations of $\kappa_2$ for the two parameter sets. The lattice QCD results are taken from Ref.~\cite{HotQCD:2018pds}.}
    \label{tab:tablek2}
\end{table}
%---------------------------------------------------------------------
We,  first, summarize the results obtained without the vector interactions in Table~\ref{tab:tablek2}. We would particularly like to draw attention to $\kappa_{2}^{B,\mu_{S}=0}$ and $\kappa_{2}^{B, n_{s}=0}$. These two curvature coefficients are almost independent of the model parameter sets and have very good agreement with the lattice estimations. On the other hand, $\kappa_{2}^{S}(\mu_{B}=0)$ is sensitive to the choice of model parameters, especially the 't Hooft determinant, $K$. Due to this robust nature of $\kappa_{2}^{B,\mu_{S}=0}$ and $\kappa_{2}^{B, n_{s}=0}$, we will use them to constrain the strength of the vector interactions and will simply study the effect of these interactions on $\kappa_{2}^{S,\mu_{B}=0}$. The LQCD data, with which the ranges of $G_V$ or $g_V$ are constrained, have improved significantly in the last few years~\cite{Kaczmarek:2011zz,Bellwied:2015rza,Bonati:2015bha,HotQCD:2018pds,Borsanyi:2020fev}. In light of the latest LQCD data, we vary $G_V$ and $g_V$ in units of $G_S$, respectively, and examine the corresponding variations in $\kappa_2^B$ and $\kappa_4^B$. In Fig.~\ref{fig:kappa2_v}, we illustrate the dependence of $\kappa_2^B$ on $G_V/G_S$ and $g_V/G_S$ for Model-\RNum{1} and Model-\RNum{2}, respectively. Compared to Refs.~\cite{Kashiwa:2011td,Bratovic:2012qs, Contrera:2012wj, Friesen:2014mha, Steinheimer:2014kka, Contrera:2016rqj}, which consider only flavor independent vector interaction, we explore both the flavor-independent and flavor-dependent interactions. We also do not include the background gauge field, intending to investigate the effects of the vector interactions on the chiral crossover line solely using the chiral dynamics, particularly because the NJL model is known to capture these dynamics reliably.

It is important to emphasize that a negative value of the vector coupling increases the effective chemical potential of the $i$th flavor ($\tilde\mu_i$) [see Eq.~\eqref{eq:effective_mu}], leading to an increase in the corresponding number density, implying the interaction to be attractive in nature. Similarly, a positive value of the same implies a repulsive-type interaction. Needless to say, the phase diagram in the $T-\tilde\mu_i$ plane is independent of the strength of the vector interaction. Hence, for attractive interaction, the bare chemical potential ($\mu_{i}$) is less than the effective chemical potential ($\tilde\mu_{i}$). This results in a faster decrease of $T_{pc}$ in the $T-\mu_{i}$ plane with increasing $\mu_{i}$ leading to an increase in $\kappa_{2}^{i}$. Conversely, for repulsive interaction, $\mu_{i}$ is greater than $\tilde\mu_{i}$ resulting in slower decrease of $T_{pc}$ leading to a decrease in $\kappa_{2}^{i}$. This shift in chemical potential, $\mu_{i}$, accounts for the observed monotonic decrease in $\kappa_2^{i}$ as the repulsive strength increases. Following Eq.~\eqref{eq:chemicalpotentials} one can relate $\kappa_2^B$ of the $T-\mu_{B}$ plane with $\kappa_{2}^{i}$ of the $T-\mu_{i}$ plane by a factor of $9$. 

%---------------------------------------------------------------------
\begin{table}[h!]
    \centering
    \begin{tabular}{|c|c|c|}
        \hline
         & ~$\kappa_2^{B,\mu_S = 0}$~  & $\kappa_2^{B, n_s=0}$ \\
        \hline
         $G_{V}/G_S$ & $\left[-0.117, 0.212\right]$ & $\left[-0.038, 0.334 \right]$ \\
        \hline
         $g_{V}/G_S$ & $\left[-0.296, 0.552\right]$  & $\left[-0.076, 0.667\right]$ \\        \hline
    \end{tabular}
    \caption{Range of $G_V/G_S$ and $g_V/G_S$ constrained from the variation of $\kappa_2^{B}$ under different conditions with parameter Set-\RNum{1}.}
    \label{tab:table_GgV}
\end{table}
%----------------------------------------------------------------------
With the aforementioned observations, we have further constrained the ranges of $G_V/G_S$ and $g_V/G_S$ for Models \RNum{1} and \RNum{2} using lattice QCD data. In Fig.~\ref{fig:kappa2_v}, we have plotted the lattice QCD results for $\kappa_2^B$, obtained from the HotQCD Collaboration~\cite{HotQCD:2018pds}. While similar studies have been conducted by the WB collaboration~\cite{Borsanyi:2020fev}, the HotQCD results are particularly relevant, as they provide data for both the $\mu_S = 0$ case and the strangeness neutrality condition ($n_s = 0$). These results are represented by the blue and coral bands, respectively, in Fig.~\ref{fig:kappa2_v}.

The points where our estimates intersect with the LQCD bands determine the allowed upper and lower bounds of the vector interaction strengths. The resulting ranges of $G_V/G_S$ and $g_V/G_S$ are summarized in Table~\ref{tab:table_GgV}. To understand the effect of model parameters on the estimated range of vector coupling strength, we have used two distinct parameter sets as given in Table~\ref {tab:parameter}. From the upper panel of Fig.~\ref{fig:kappa2_v}, it is evident that the allowed ranges of the vector coupling strength for the two parameter sets are very close to each other, while being constrained by the lattice QCD estimation of $\kappa_2^B$. It appears that such models exhibit some characteristic features in this regard and the values of $\kappa_2^B$ are largely independent of the parameters involved~\cite{Ali:2024xxx}. Hence, in the following discussions, we will be presenting all the findings only for parameter set \RNum{1}.

We find the allowed ranges of the vector interactions accommodate both attractive and repulsive interactions for both models. The estimated range of $G_V$ for Model-\RNum{1} is in line with the findings of Refs.~\cite{Kashiwa:2011td,Bratovic:2012qs, Contrera:2012wj, Friesen:2014mha, Steinheimer:2014kka, Contrera:2016rqj}, however, most of which favored a stronger repulsive vector interaction as compared to the present results. This is predominantly due to the available LQCD data which have evolved significantly over time and settled for  comparatively higher values~\cite{Kaczmarek:2011zz,Bellwied:2015rza,Bonati:2015bha,HotQCD:2018pds,Borsanyi:2020fev}.

Although Model-\RNum{1} and Model-\RNum{2} are two different models, in certain limits, one can establish a connection between $G_V$ and $g_V$. For example, with $\mu_S = 0$, one expects $g_V = 3.0 G_V$, assuming degenerate masses for the light and strange quarks leading to $n_u=n_d=n_s$. However, in the scenario considered here, $g_V$ is less than $3 G_V$ due to the much higher mass of the strange quark compared to the light quarks. In contrast, under the strangeness neutrality condition ($n_s=0$), the relation becomes $g_V = 2 G_V$ as isospin symmetry ensures $n_u=n_d$. The coupling strength ranges derived in Table~\ref{tab:table_GgV} follow these conditions. 

For small $\mu_B/T$ limit, the parametric form of Eq.~\eqref{eq:curvature} has a weaker dependence on $\kappa_4^B$. The estimations from Lattice QCD are zero within the variances, and we have also found a similar trend from our effective model analysis. For a wider range of coupling strength, the $\kappa_4^B$ lies near zero for both models, providing no additional improvement over the constraint from $\kappa_2^B$. Such variations have been presented in Appendix~\ref{sec:app}.

We wish to reiterate that in Model-\RNum{1}, there is a noticeable crossing between the cases of $\mu_S = 0$ and $n_s = 0$ for both parameter sets, whereas such a feature is absent in Model-\RNum{2}, where an almost parallel trend seems to exist. In Ref.~\cite{Ali:2024nrz},  the authors demonstrated that a finite value of $\mu_S$ tends to decrease $\kappa_2^B$, which is primarily responsible for the lower value of $\kappa_2^B$ when considering the $n_s = 0$ case. According to the analytical expression for the effective chemical potential in Model-\RNum{2}, as given in Eq.~\eqref{eq:chemicalpotentials}, the strangeness neutrality condition explicitly requires $\mu_S = \mu_B/3$. This introduces a positive shift in $\mu_S$ relative to the $\mu_S = 0$ case, leading to the observed parallel behavior and the absence of any crossing within the explored range of the coupling strength $g_V$. In contrast, for Model-\RNum{1}, the $n_s = 0$ condition does not fix $\mu_S$ to a specific value but instead makes it dependent on the light quark number densities as $\mu_S = \mu_B/3 - 2G_{V}(n_u + n_d)$. This structure implies that at a certain positive $G_V$, $\mu_S$ will reach zero, explaining the crossing with the $\mu_S = 0$ case in Model-\RNum{1}. 

As pointed out in Table~\ref{tab:tablek2}, the curvature coefficient $\kappa_{2}^{S,\mu_{B}=0}$ is strongly influenced by the model parameters, especially by the 't Hooft determinant term, $K$. In the absence of a vector interaction, the effect of the strange quark sector propagates to the light quark sector through the mixing term, proportional to $K$ as given in Eq.~\eqref{eq:mass}. Without the 't Hooft interaction, the pseudo-critical temperature $T_{pc}$ would be independent of $\mu_{S}$ resulting in $\kappa_{2}^{S,\mu_{B}=0}=0$. Even though the light quark chemical potential gets modified through $n_s$ at finite $\mu_{S}$, the major contributions arise through the flavor mixing in the constituent masses. As a result, we observe that $k_{2}^{S,\mu_{B}=0}$ has a weak dependence on the vector interaction, which is further influenced by the choice of the parameter set and thus cannot be used to obtain any further improvement on the bounds found from $k_{2}^{B,\mu_{S}=0}$ and $k_{2}^{B,n_{s}=0}$. For better visual understanding, we have provided the nature of $\kappa_{2}^{S}$ as a function of $G_{V}/g_{V}$ in Fig.~\ref{fig:kappa2_S} of the Appendix. 

At this point, it is worth noting that a similar variation of $\kappa_{2}^{B,\mu_{S}=0}$ with varying strength of vector interaction (as shown in Fig.\ref{fig:kappa2_v}) had earlier been explored in Polyakov loop extended NJL (PNJL) model ~\cite{Bratovic:2012qs, Contrera:2012wj, Friesen:2014mha, Contrera:2016rqj}. These studies further employed the lattice QCD estimation of $\kappa_{2}^{B,\mu_{S}=0}$, available at the time, to constrain the range of those vector interactions. Our results exhibit a similar behavior and are in line with those from PNJL. Although the constraints values $G_V$ and $g_V$ from these PNJL studies favored a positive range. This is primarily because the lattice results available at that time were significantly lower in values than the present estimates.

%%%%%%%%%%%%%%%%%%%%%%%%%%%%%%%%%%%%%%%%%%%%%%%%%%%%%%%%%%%%%%%%%%%%%%%%%%%%%%%%%%%%%%%%%%%%%%%%%%%
\subsection{$\mu_S$ dependence of $\kappa_2^{B}$} 
\label{ssec:kappa_vs_mus}
%%%%%%%%%%%%%%%%%%%%%%%%%%%%%%%%%%%%%%%%%%%%%%%%%%%%%%%%%%%%%%%%%%%%%%%%%%%%%%%%%%%%%%%%%%%%%%%%%%%
%-----------------------------------------------------------------------------------
\begin{figure}[h]
\begin{center}
  \includegraphics[scale=0.56]{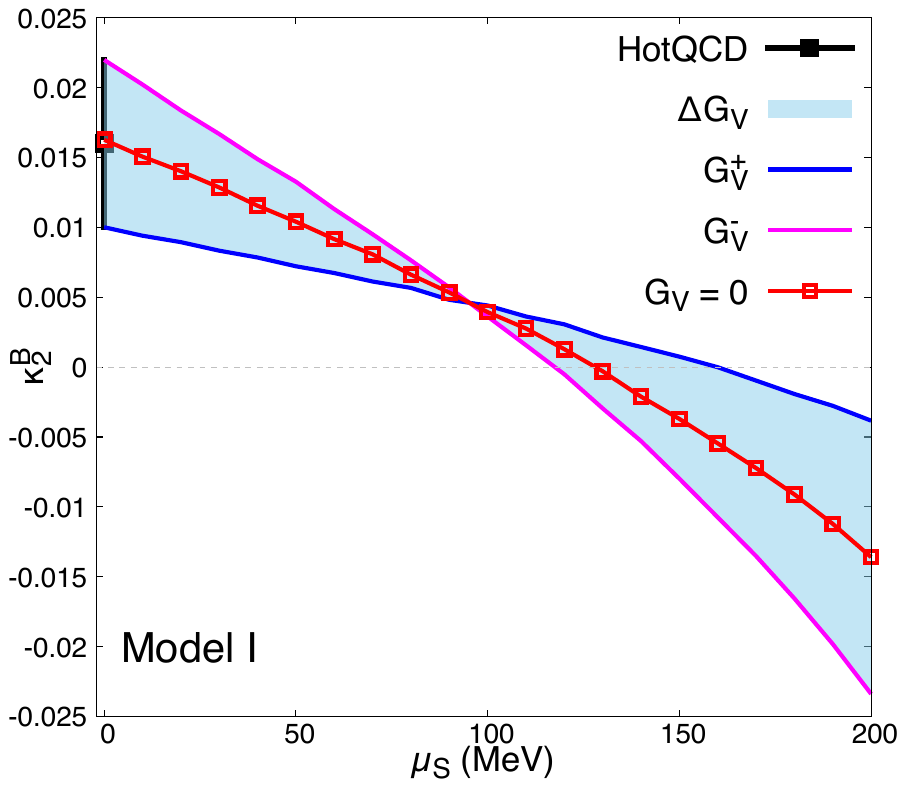}
  \includegraphics[scale=0.56]{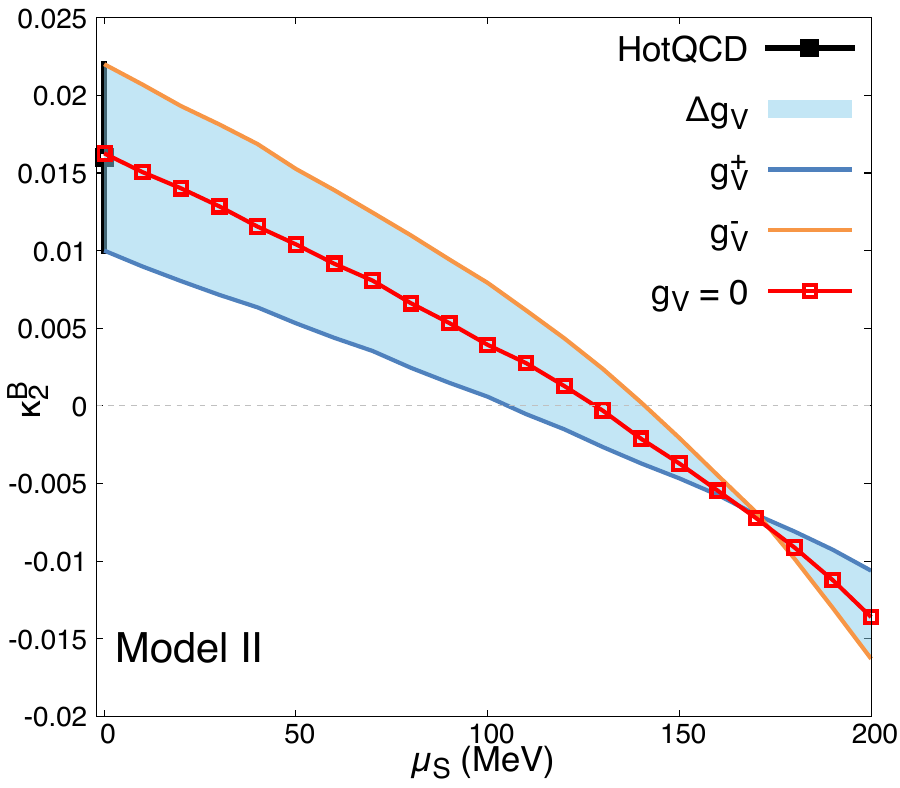}
  \caption{The curvature coefficient ($\kappa_2^B$) as a function of strangeness chemical potential ($\mu_S$). $[G_V^-, G_V^+]$ and $[g_V^-, g_V^+]$ are the allowed ranges of vector interaction found using LQCD data. Red data points are without the vector interaction i.e., $G_V=g_V=0$. Black point represents the LQCD estimations of the $\kappa^{B}_{2}$ at $\mu_S=0$~\cite{HotQCD:2018pds}.}
  \label{fig:kappa_vs_mus}
\end{center}  
\end{figure}
%------------------------------------------------------------------------------------
In a recent work~\cite{Ali:2024nrz}, some of us investigated the dependence of the curvature coefficients on $\mu_S$ and proposed a novel approach for determining the strength of the 't Hooft determinant interaction, $K$. The 't Hooft term induces mixing between light and heavy quark flavors, making the precise determination of its strength critical for estimating the degree of flavor mixing in a $2+1$ NJL model. On the other hand, in a $2$-flavor NJL model, such mixing effect between the two light flavors gets reflected only in the isospin asymmetric scenarios, as shown in Ref.~\cite{Ali:2020jsy, Ali:2021zsh}. In the presence of a flavor-independent vector interaction, the effective chemical potential, $\tilde\mu_i$, depends on the densities of all quark flavors. As a result, the term $G_V \sum_i n_i$ introduces a form of flavor mixing along with the 't Hooft interaction induced mixing through the constituent mass [see Eq.~\eqref{eq:mass}].

To quantify this, we analyzed the variation of $\kappa_2^B$ as a function of $\mu_S$. The results are presented in Fig.~\ref{fig:kappa_vs_mus}. The general trend shows that as $\mu_S$ increases, $\kappa_2^B$ decreases and eventually becomes negative at higher values of $\mu_S$ commensurate with the observations of Ref.~\cite{Ali:2024nrz}. We provide estimates constrained by the vector interaction strengths for both Model-\RNum{1} (upper panel, Fig.~\ref{fig:kappa_vs_mus}) and Model-\RNum{2} (lower panel), represented by the cyan bands. 

One notable feature observed is the variation of $\kappa_2^B$ with increasing $\mu_S$, where the upper and lower curves, corresponding to lowest and highest allowed values of $G_{V}/g_{V}$, respectively, cross each other, intersecting the $G_V=0$ line around the same region. In Model-\RNum{1}, this crossing can be interpreted as follows: at the crossing point, the effect of $G_V$ becomes negligible. For positive and sufficiently large $\mu_S$ compared to $\mu_{B}/3$, the net strange quark density can become negative, potentially offsetting the net light quark density. With a small total quark number density (very close to zero), at low $\mu_{B}$'s, the light quark chemical potential gets a very small modification due to vector interaction, as can be seen from Eq.~\eqref{eq:effective_mu}. At this point, the results should coincide with those for $G_V=0$, leading to the observed crossing of the two curves (magenta and blue).

%~~~~~~~~~~~~~~~~~~~~~~~~~~~~~~~~~~~
\begin{figure}[h!]
\subfloat{\includegraphics[scale=0.56]{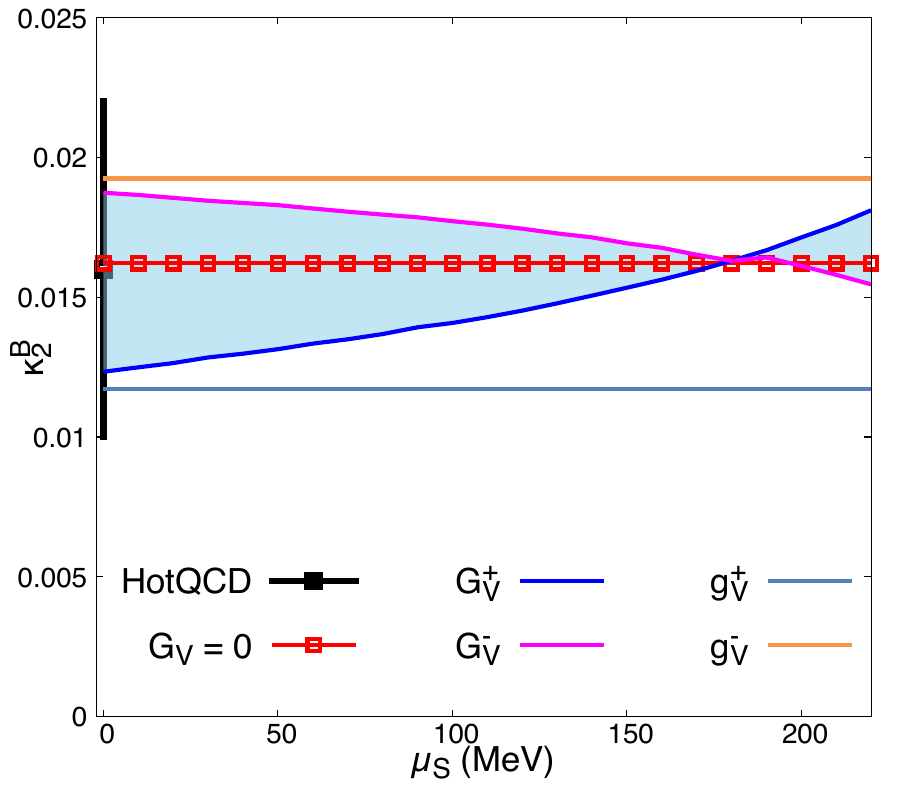}}
%\subfloat{\includegraphics[scale=0.56]{Fig-k4_M1Gd0.pdf}}
\caption{Curvature coefficient, $\kappa_2^B$ as a function of $\mu_{S}$ with $K=0$ for both Model-\RNum{1} and Model-\RNum{2}.}
\label{fig:k2_K0}
\end{figure}
%~~~~~~~~~~~~~~~~~~~~~~~~~~~~~~~~~~~
At this juncture, we want to separately assess the impact of flavor mixing. In Ref.~\cite{Ali:2024nrz}, it has been observed that this decreasing trend of $\kappa_2^B$ is due to the flavor mixing between the light and strange quark sectors. The flavor mixing primarily happens for the 't Hooft interaction term, represented by the $K$, whereas the presence of the vector interaction also provides scope for the flavor mixing, especially in the model \RNum{1}, where the vector interaction is flavor-independent. To separate out the effect arising only from $G_V$, we have estimated the $\kappa_2^B$ variation with $\mu_S$ in the case of $K = 0$ in Fig.~\ref{fig:k2_K0}. As expected, the $\kappa_2^B$ does not vary with the $\mu_S$ for $G_V=0$, as there is no flavor interdependency from both $K$ and $G_V$. Whereas for a finite $G_V$, the introduction of a positive strangeness chemical potential $\mu_S$, increases the abundance of anti-strange quarks in the system for a fixed baryon chemical potential, $\mu_B$. In contrast, the number densities of up and down quarks remain positive as they are only dependent on $\mu_B$. As a result, increasing $\mu_S$ reduces the total number density, thereby diminishing the effect of the vector interaction [see Eq.~\eqref{eq:effective_mu}]. This leads to a cancellation of the vector interaction effect, $G_V$, at a specific value of $\mu_S$, resulting in this crossing.

It is important to note that the crossing points differ between Model-\RNum{1} and Model-\RNum{2}. In Model-\RNum{2}, this crossing corresponds to a negative value of $\kappa_2^{B}$ and occurs at a higher value of $\mu_S$ (Fig.~\ref{fig:kappa_vs_mus}). This suggests that $\mu_S$ has a weaker influence on the chiral transition line in the presence of flavor-dependent vector interactions $(g_V\ne0)$. Moreover, for $K = 0$, the flavor-dependent interaction related to $\mu_S$ does not impact the light quark chemical potential, leading to a curvature coefficient that remains independent of $\mu_S$. This is demonstrated by the two straight lines, representing $g_V^+$ (soft blue) and $g_V^-$ (soft orange), in Fig.~\ref{fig:k2_K0}. This indicates that in Model-\RNum{2}, there is no flavor mixing due to nonzero $g_V$ and the reduction in $\kappa_2^B$ with increasing $\mu_S$, as presented in the lower panel Fig.~\ref{fig:kappa_vs_mus}, is entirely due to the flavor mixing term $K$.

We note that a recent lattice QCD study~\cite{Ding:2024sux} has investigated the chiral phase transition as a function of both $\mu_{B}$ and $\mu_{S}$, and determined the leading-order curvature coefficients. From their parametric representation of the phase line, one can extract $\kappa_{2}^{B}$ for chosen values of $\mu_{S}$. Interestingly, this parametrization exhibits the same decreasing trend of $\kappa_{2}^{B}$ with increasing $\mu_{S}$, with $\kappa_{2}^{B}$ turning zero around $\mu_{S}\simeq 120~\text{MeV}$. The estimations of higher-order curvature coefficients from lattice QCD at larger $\mu_{S}$ would allow a more quantitative assessment of the strength of the ’t~Hooft and vector interactions.

%%%%%%%%%%%%%%%%%%%%%%%%%%%%%%%%%%%%%%%%%%%%%%%%%%%%%%%%%%%%%%%%%%%%%%%%%%%%%%%%%%%%%%%%%%%%%%%%%%%
\subsection{$\mu_S$ variation of $\kappa_4^{B}$} 
\label{ssec:kappa_vs_mus}
%%%%%%%%%%%%%%%%%%%%%%%%%%%%%%%%%%%%%%%%%%%%%%%%%%%%%%%%%%%%%%%%%%%%%%%%%%%%%%%%%%%%%%%%%%%%%%%%%%%
\begin{figure}[h!]
\begin{center}
  \includegraphics[scale=0.56]{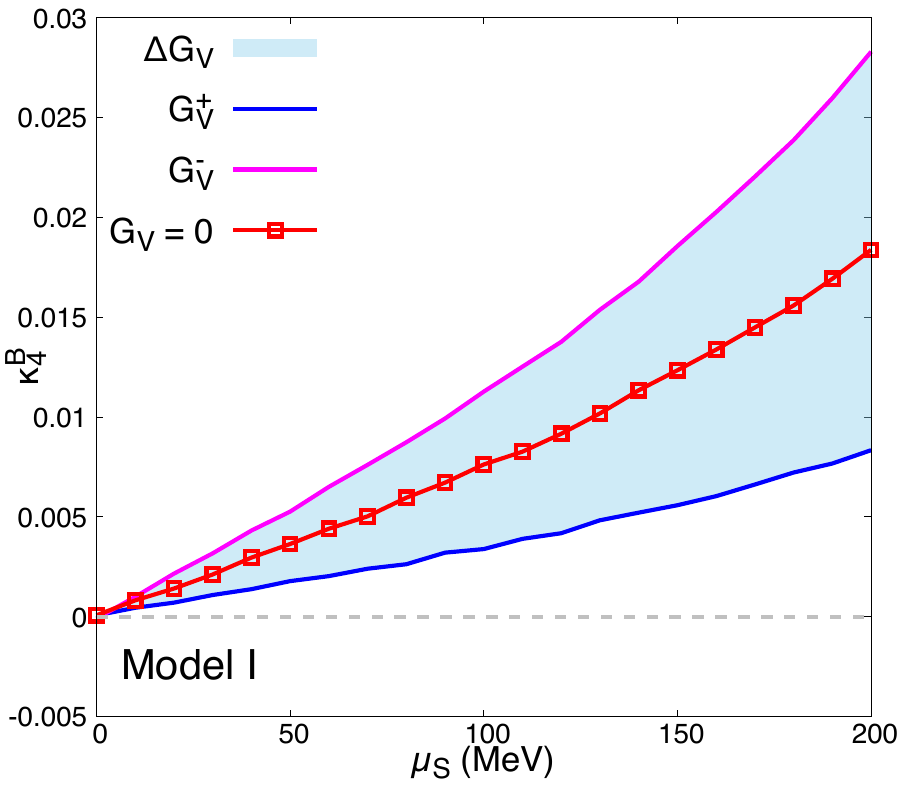}
  \includegraphics[scale=0.56]{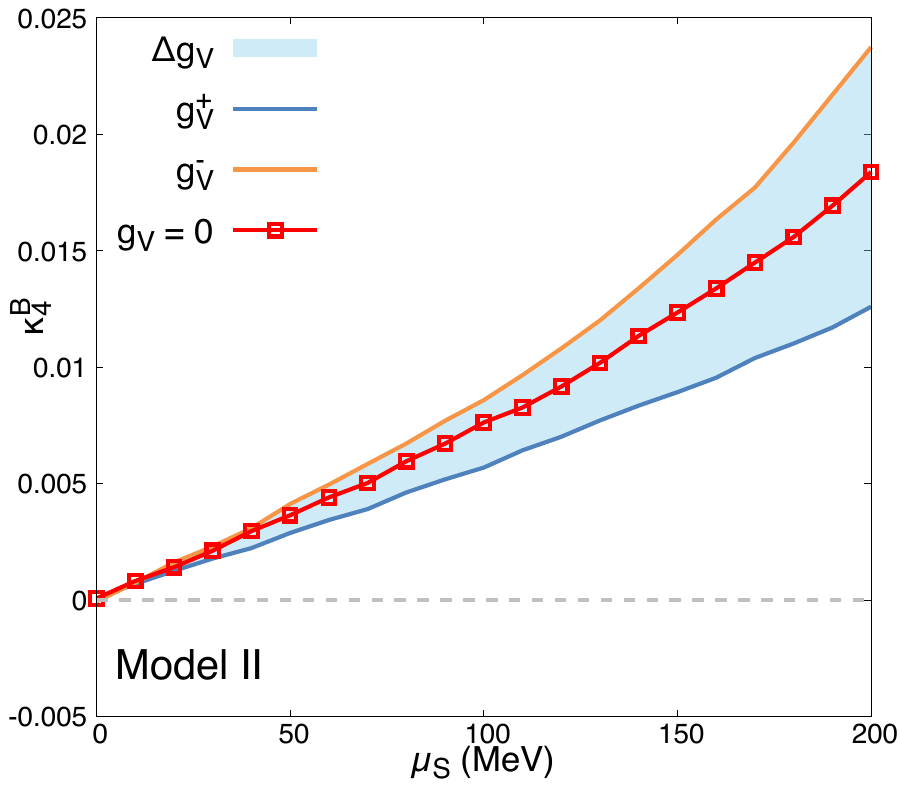}
  \caption{The curvature coefficient for different values of $\mu_S$ in Model-II with the same parameter sets used in the above graph.}
  \label{fig:kappa_vs_musM2}
\end{center}  
\end{figure}
Although, in scenarios considered in LQCD, the curvature coefficient $\kappa_{4}^{B}$ are obtained to be zero, we have obtained its significant dependence on $\mu_{S}$. Fig.~\ref{fig:kappa_vs_musM2} represents the models' predictions on $\kappa_{4}^{B}$ as a function of $\mu_{S}$. The overall behaviour of $\kappa_{4}^{B}$ as a function of $\mu_S$ is the same (increasing) for both Model-\RNum{1} and Model-\RNum{2} with the corresponding attractive interactions $(G_V^{-}/g_V^{-})$ producing a higher curvature as compared to repulsive ones for a given value of $\mu_S$.  This observation might provide us with a better understanding of the effect of $\mu_{S}$ as well as the vector interaction on the crossover line once explored in LQCD.

%%%%%%%%%%%%%%%%%%%%%%%%%%%%%%%%%%%%%%%%%%%%%%%%%%%%%%%%%%%%%%%%%%%%%%%%%%%%%%%%%%%%%%%%%%%%%%%%%%%%
\subsection{Effect of vector interaction on the critical endpoint}
\label{ssec:res_cep}
%%%%%%%%%%%%%%%%%%%%%%%%%%%%%%%%%%%%%%%%%%%%%%%%%%%%%%%%%%%%%%%%%%%%%%%%%%%%%%%%%%%%%%%%%%%%%%%%%%%%
%---------------------------------------------------------------------------------
\begin{figure}[h!]
\begin{center}
  \includegraphics[scale=0.56]{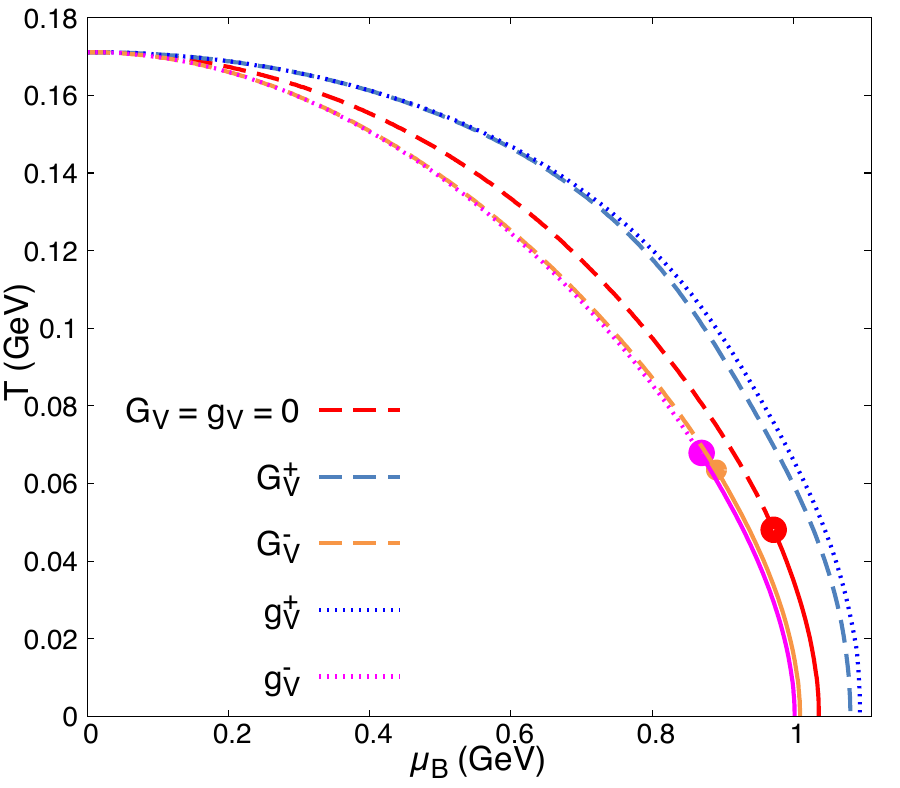}
  \caption{Effect of the vector interactions on the QCD phase line for $\mu_S = 0$. Blue and magenta lines denote the estimation from the upper and lower bounds of the flavor-dependent vector interaction $g_V$, respectively. Soft blue and soft orange denote the same for flavor-independent scenario, i.e., with nonzero $G_V$. The red line denotes the limiting case with no vector interaction, $g_V = G_V = 0$.}
  \label{fig:PDL}
\end{center}  
\end{figure}
%-----------------------------------------------------------------------------------
For physical light quark masses, the QCD transition is expected to be a crossover at zero or low baryon densities. However, at high baryon densities the transition is expected to be first order. Thus, the crossover is expected to meet the first order phase transition at some critical endpoint (CEP). It is important to investigate the impact of constrained vector interaction strengths on the location of this CEP. Previous studies have shown that the CEP's position shifts based on the strength of the vector interaction~\cite{Fukushima:2008wg, Friesen:2014mha, Sun:2020bbn}. Non-zero couplings, $G_V$, and $g_V$, both affect the curvature coefficients and the extension of the pseudo-critical (cross-over) region in the phase diagram.

In Fig.~\ref{fig:PDL}, we show the evaluated phase lines for constrained values of $G_V$ and $g_V$, with the corresponding CEPs indicated by circle symbols. The exact locations of the CEPs are provided in Table~\ref{tab:table_CEP}. As the repulsive vector interaction flattens the phase boundary, the CEP is expected to shift rightward for the positive upper limits of both $G_V$ and $g_V$, resulting in a corresponding decrease in $T^{\text{CEP}}$, as seen for the $g_V^+$ line. It is important to note that in Model-\RNum{1}, all three quark flavors contribute to the shifted chemical potential $\tilde{\mu}$. In contrast, Model-\RNum{2} includes flavor-dependent contributions, where only individual quark flavors affect $\tilde{\mu}$. As a result, the phase boundary is more compressed in the flavor-independent scenario. These results align with those in Ref.~\cite{Fukushima:2008wg}. For the negative lower values of $G_V$ and $g_V$, the CEP shifts leftward, with an associated increase in $T^{\text{CEP}}$. It should be emphasized that though the locations of these CEPs are different from those found using functional methods~\cite{Gunkel:2021oya,Gao:2020fbl}, none of them contradicts predictions from Lattice QCD~\cite{Borsanyi:2021hbk,Philipsen:2021qji}, and they offer a baseline within this effective model framework. However, another important ingredient of the model, the strength of the 't Hooft determinant term $(K)$, plays crucial role on determining the location of the CEP~\cite{Fukushima:2008wg}. At present, there are some arbitrariness for the value of $K$ (Table~\ref{tab:parameter}). A precise determination of $K$, possibly by following the technique mentioned in Ref.~\cite{Ali:2024nrz}, along with future constraints on the CEP location from first-principle QCD calculations, may help narrow down the range of the vector interaction strengths explored here.

%---------------------------------------------------------------------
\begin{table}[h!]
    \centering
    \begin{tabular}{|c|c|c|}
        \hline
         $G_V, g_V$ & $\mu^{CEP}_B$ (GeV) & $T^{CEP}$ (GeV)\\
        \hline
          $G_{V} = g_{V} = 0 $  & 0.972  & 0.0481  \\
        \hline  
          $G^{-}_{V}$  & 0.891  & 0.0635  \\
        \hline    
          $G^{+}_{V}$  & NA   & NA  \\
        \hline    
          $g^{-}_{V}$  & 0.870  & 0.0679  \\
        \hline
          $g^{+}_{V}$  & NA  & NA    \\
        \hline    
    \end{tabular}
    \caption{Locations of the CEP for different values of $G_V$ and $g_V$.}
    \label{tab:table_CEP}
\end{table}
%----------------------------------------------------------------------

We want to mention that the obtained locations of the CEP with the other parameter set (set~II) are consistent with earlier works ~\cite {Gastineau:2001zke, Costa:2007ie, Costa:2008yh}. The effect of vector interactions on the CEP locations is qualitatively similar to that of set~I. As $T^{\rm CEP}$ from set~II is significantly larger compared to the one from parameter set~I, this necessitates a stronger repulsive interaction to make the CEP disappear from the phase diagram. As the allowed range of $G_V$ in our work is almost similar for both parameter sets, the CEP does not disappear from the phase diagram for set~II even for the repulsive interaction. It should be emphasized that the predictions of CEP locations within the NJL-like model are very sensitive to the model parameters~\cite{Bhattacharyya:2010wp}. For example, stronger values of the eight-quark coupling $g_1$ would push the temperatures associated with CEP to even larger values~\cite{Hiller:2008nu}. In the present study, we do not intend to quantify the absolute location of the CEPs; rather, we focus on their shifts with varying vector interaction strength.

%%%%%%%%%%%%%%%%%%%%%%%%%%%%%%%%%%%%%%%%%%%%%%%%%%%%%%%%%%%%%%%%%%%%%%%%%%%%%%%%%%%%%%
\section{Summary and Outlook}
\label{sec:sum}
%%%%%%%%%%%%%%%%%%%%%%%%%%%%%%%%%%%%%%%%%%%%%%%%%%%%%%%%%%%%%%%%%%%%%%%%%%%%%%%%%%%%%%

In this work, we investigate the effect of vector-type interactions within the $2+1$-flavor Nambu\textendash Jona-Lasinio (NJL) model, considering both flavor-independent and flavor-dependent interactions with coupling strengths $G_V$ and $g_V$, respectively. In general, the strengths of these coupling constants can be fixed from the masses of the vector mesons~\cite{Vogl:1991qt,Klevansky:1992qe}. However, in an effective model scenario, such an exercise is less robust than the determination of the scalar coupling constant. On the other hand, in a dense medium, vector interactions can be induced due to their direct coupling with the number density operator~\cite{Fukushima:2008wg,Abuki:2009ba}. Such induced vector interactions arise only in a finite-density regime, making it difficult to fix the strengths of induced $G_V$ and $g_V$ from the QCD vacuum properties. Previous attempts to constrain these coupling strengths using LQCD-estimated curvature coefficients of the crossover line were limited due to large uncertainties in the lattice results. 

Here, we revisit this issue and utilize the precise lattice results for $\kappa_2^{B}$ for the cases $\mu_S=0$ and $n_s=0$ to constrain both $G_V$ and $g_V$. We varied these couplings in units of $G_S$ over a range of negative to positive values and evaluated $\kappa_{2,4}^B$ for the chiral crossover line. Negative values of $G_V$ and $g_V$ produce larger values of $\kappa_2^{B}$, which decrease as $G_V/G_S$ and $g_V/G_S$ increase towards positive values and further. This approach provides a tighter constraint on the range of these interaction strengths. In the $\mu_S = 0$ case, the allowed ranges of interaction strength are $\left[ -0.117, 0.212 \right]$ and $\left[ -0.296, 0.552 \right]$ (in units of $G_S$) for the flavor-independent and flavor-dependent interactions, respectively. Conversely, the $n_s = 0$ case provides narrower ranges: $\left[ -0.038, 0.334 \right]$ for $G_V/G_S$ and $\left[ -0.076, 0.667 \right]$ for $g_V/G_S$. As the estimates for $\kappa_4^B$ are consistent with lattice results and remain within variances for a wide range of $G_V, g_V$, they do not further constrain the vector interaction strength. At this point, we want to iterate that the bound on $G_V, g_V$ we get for the scenarios $\mu_S=0$ and $n_s=0$ are different, which necessitates the consideration of a more realistic interaction strength which depends on $\mu$.

We find that the improved lattice data on $\kappa_{2}^{B}$ provides bounds on the vector interactions accommodating both attractive and repulsive vector interaction within the model framework. Such findings highlight the fact that lattice-calculated bounds on curvature coefficients, which has been previously used to constrain induced vector interactions~\cite{Bratovic:2012qs, Contrera:2012wj, Friesen:2014mha, Contrera:2016rqj,Steinheimer:2014kka}, allow for positive and negative induced vector interactions. However, such a bound should be tested against other phenomenological consequences of the vector interaction; for instance, neutron star observations favor a repulsive (positive) vector interaction, which is required to generate an equation of state stiff enough to support two-solar-mass stars~\cite{Kojo:2014rca,Masuda:2012kf}. On the other hand, attractive vector interaction can lead to larger values of the susceptibilities and thus mechanical instability at high density~\cite{Islam:2014sea}, particularly with a constant vector interaction. Such uncertainties make the strength and sign of the vector interaction a critical issue. Future lattice QCD determinations of susceptibilities at larger baryon densities will therefore be invaluable in constraining the magnitude of ${G_V} ~\text{and}~ {g_V}$.

It should also be mentioned that the masses of vector mesons allow only repulsive interactions~\cite{Vogl:1991qt}. However, one needs to take into account both the quark loop and the pseudoscalar meson loops in the calculation of vector meson self-energy~\cite{Vogl:1991qt}; otherwise, the vector interaction may appear with the wrong sign~\cite{Sheng:2022ssp}. Using vector meson masses, Ref.\cite{Vogl:1991qt} extracted a value of $G_V/G_S \sim 1.5$, whereas Ref.\cite{Abuki:2009ba} predicts the ratio to be $\sim 0.6$, thus allowing for a wide range depending on the model parameters. The upper limits of the bounds on the induced vector interaction that we find are therefore not unreasonable with such a wide range, if one attempts to determine the vector meson masses from an induced vector interaction.

While testing the model's predictions under the zero strangeness or strangeness neutrality condition, we observe that $\kappa_{2}^{B}$ is always smaller in the latter case for both increasing $G_V$ and $g_V$, except at higher values of $G_V$ (flavor-independent scenario), where we observe a crossing. However, the bound is always wider on $g_V$ as compared to $G_V$, which can be explained within the model framework. The models' predictions on the fourth-order curvature coefficient $(k_4^B)$ match with the corresponding LQCD results, which can be termed as consistent with zero with the error bars. We intentionally avoided the Polyakov loop dynamics to focus solely on the chiral dynamics. However, a similar analysis has been performed in the presence of Polyakov loop dynamics in Refs.~\cite{Bratovic:2012qs, Contrera:2012wj, Friesen:2014mha, Contrera:2016rqj}, where the curvature coefficients $\kappa_{2}^{B}$ were found to be very close to our NJL estimate. Since the curvature is calculated for the transition line up to $\mu_{B}/T \leq 1$, we expect the allowed range of the vector interaction strength not to change significantly in the presence of a background gauge field. Nevertheless, this is certainly an interesting aspect to explore and verify explicitly, which we plan to take up in future work.

In the NJL model, the mixing between light and strange quarks primarily arises from the 't Hooft determinant term, introduced to take into account the QCD axial anomaly. The flavor-independent vector interaction also couples the light and strange quark sectors through their net densities, thereby contributing to flavor mixing. In this work, we have proposed a novel method to separate out the contribution of the flavor-independent vector interaction to flavor mixing by investigating the effect of the strangeness chemical potential, $\mu_S$, on $\kappa_2^B$. These flavor mixings between light and strange sectors are responsible for the decrease in $\kappa_2^B$ as $\mu_S$ increases. Our findings are in general agreement with those from Ref.~\cite{Ali:2024nrz}. We also observe that at a specific value of $\mu_S$, the effect of the vector interactions becomes negligible for both Model-\RNum{1} and Model-\RNum{2} and match with those for $G_V=g_V=0$ limit. This value of $\mu_S$ is larger for Model-\RNum{2}, signifying weak flavor mixing in the flavor-dependent coupling as compared to the flavor-independent one. Such interesting characteristics provide us with a better understanding of the types of vector interactions and possible signatures to distinguish them.

To elaborate on the contribution from different mixing effects, we have studied the case where the 't Hooft determinant term is taken to be zero. This has enabled us to explicitly demonstrate the impact of vector interactions on the variation of $\kappa_2^B$ and allows for a direct assessment of the flavor mixing driven by this flavor-independent vector interaction. In this limit, there is no variation of $\kappa_2^B$ in model \RNum{2}, indicating no mixing effect for the flavor-dependent vector interaction.

For the final part of our study, we find the location of the CEP for the allowed ranges of vector interactions. The location of CEP is heavily impacted by the strength of the vector interaction. We find that the CEP exists for the negative lower bounds of the vector interactions and disappears for the positive upper bounds in both the Model-\RNum{1} and Model-\RNum{2}. Although the model's predictions on the location of CEP are consistent with the existing LQCD bound, one should keep in mind that such observations can, in principle, depend on multiple factors, capturing all of which remains a task beyond such simple effective model calculations.

%%%%%%%%%%%%%%%%%%%%%%%%%%%%%%%%%%%%%%%%%%%%%%%%%%%%%%%%%%%%%%%%%%%%%%%%$$
\section*{Acknowledgment} D.B. is supported by the Department of Science and Technology, Government of INDIA under the SERB National Post-Doctoral Fellowship Reference No. {\it{ PDF/2023/001762}}. C.A.I. acknowledges Goethe University’s R3 Career Support, supported by Johanna Quandt Young Academy $@$Goethe (JQYA) and the support by the State of Hesse within the Research Cluster ELEMENTS with project ID 500/10.006. He also thanks ECT* and CRC for support at the Workshop “Spin and quantum features of QCD plasma” during which this work has been completed.

\appendix
\section{Variation of $\kappa_4^B$ with $G_V$ and $g_V$}
\label{sec:app}
\begin{figure}[htb!]
\begin{center}
  \includegraphics[scale=0.56]{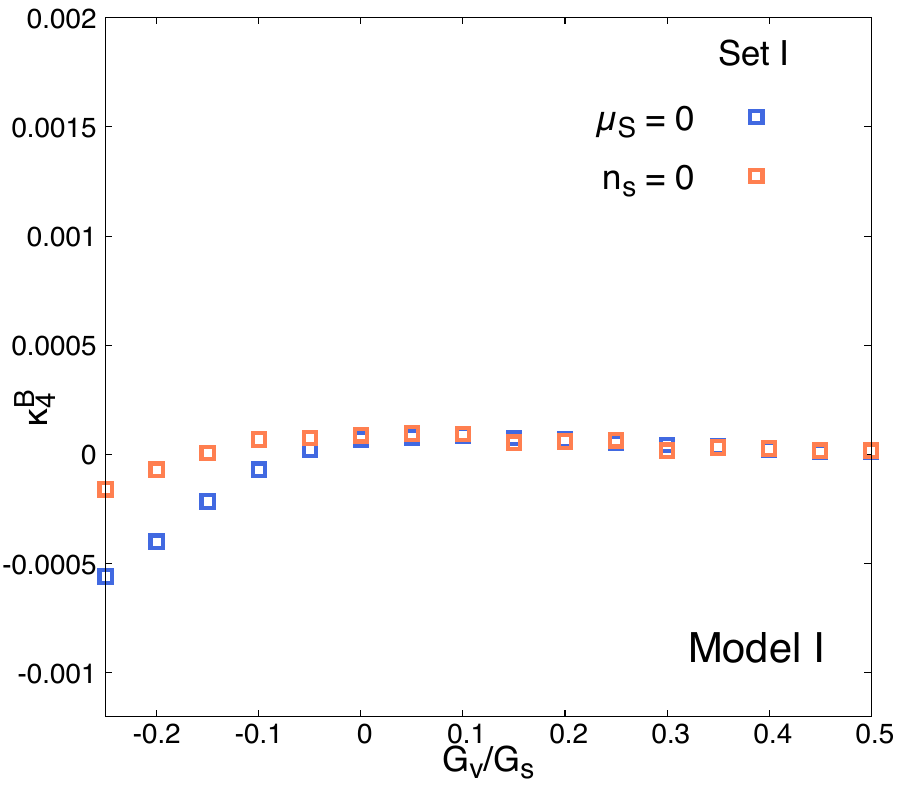}
  \includegraphics[scale=0.56]{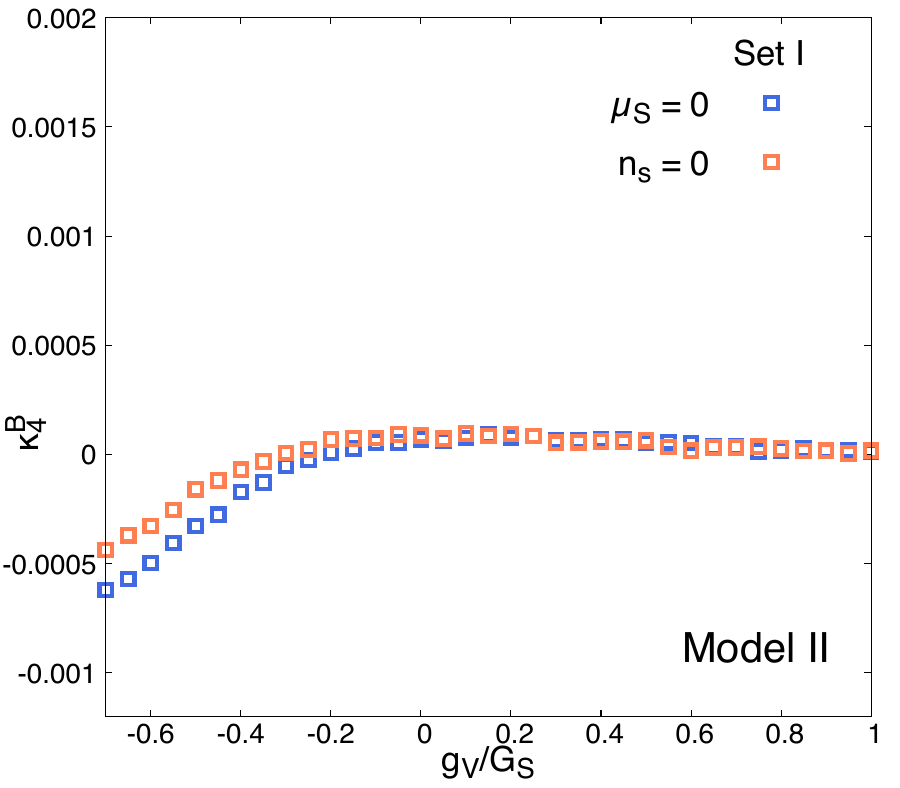}
  \caption{The curvature coefficients, $\kappa_{4}^{B}$ as a function of $G_{V}/G_{S}$ (upper panel) $g_{V}/G_{S}$ (lower panel), respectively. The LQCD estimations of the same are $0.001(7)$ and $0.004(6)$ for $\mu_{S}=0$ and $n_{s}=0$, respectively~\cite{HotQCD:2018pds}.}
  \label{fig:kappa4_v}
\end{center}  
\end{figure}

In Fig.~\ref{fig:kappa4_v}, we present the $4$th order curvature coefficient, $\kappa_{4}^{B}$, as a function of $G_{V}$ (Model-\RNum{1}) and $g_{V}$ (Model-\RNum{2}), respectively. We explored it for the same ranges of $G_{V}$ and $g_{V}$ that we used to explore both $\kappa_{2}^{B,\mu_{S}=0}$ and $\kappa_{2}^{B,n_{s}=0}$. It remains zero for a repulsive interaction (positive $G_{V}/g_{V}$) but starts to decrease as the strength of the attractive interaction (negative $G_{V}/g_{V}$) increases. The LQCD estimates of $4$th order curvature coefficient for $\mu_{S}=0$ and $n_{s}=0$ are $0.001(7)$ and $0.004(6)$, respectively \cite{HotQCD:2018pds}. Due to the large uncertainties in LQCD estimates, we cannot further constrain $G_{V}/g_{V}$ beyond the limits obtained from $\kappa_{2}^{B}$s.
%%%%%%%%%%%%%%%%%%%%%%%%%%%%%%%%%%%%%%%%%%%%%%%%%%%%%%%%%%%%%%%%%%%%%%%%$$
%%%%%%%%%%%%%%%%%%%%%%%%%%%%%%%%%%%%%%%%%%%%%%%%%%%%%%%%%%%%%%%%%%%%%%%%%%%
\section{ $\kappa_{2}^S$ as a function of $G_V$ and $g_V$}
\begin{figure}[htb!]
\begin{center}
  \includegraphics[scale=0.56]{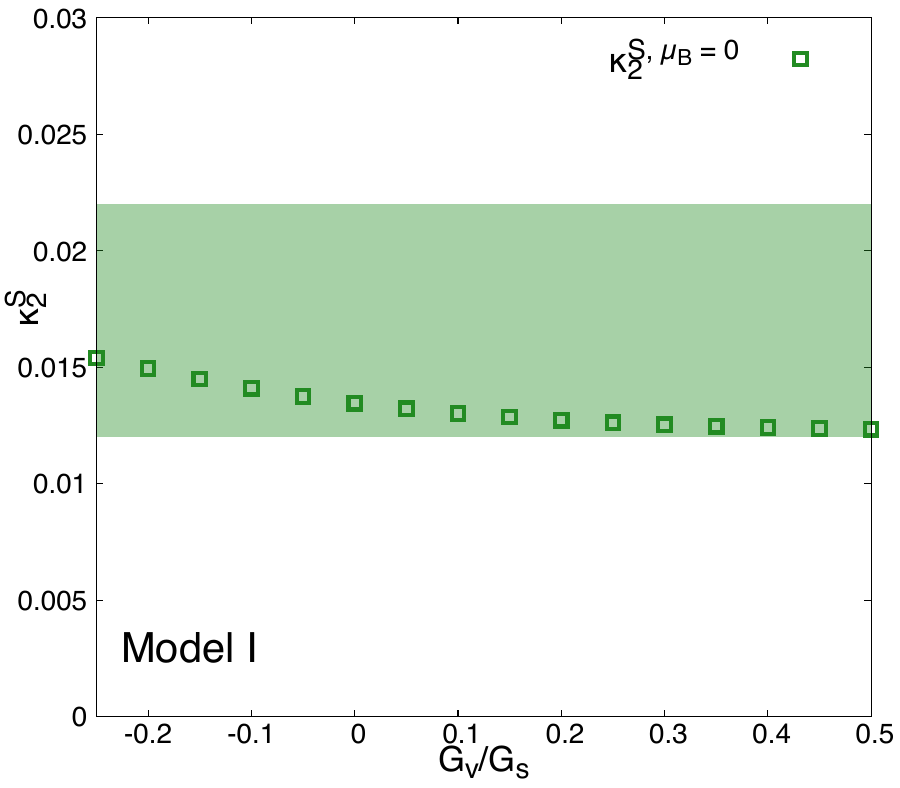}
\includegraphics[scale=0.56]{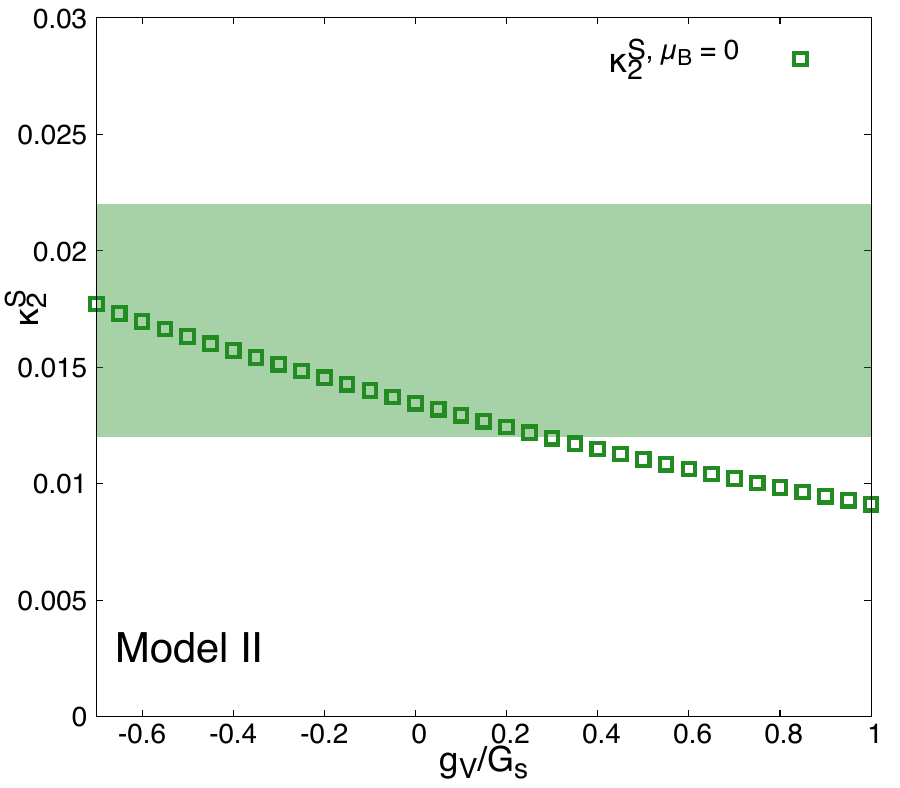}
  \caption{The curvature coefficients, $\kappa_{2}^{S,\mu_{B}=0}$ as a function of $G_{V}/G_{S}$ (upper panel) $g_{V}/G_{S}$ (lower panel), respectively. The LQCD estimation of the same is $0.017(5)$~\cite{HotQCD:2018pds}.}
  \label{fig:kappa2_S}
\end{center}  
\end{figure}

%%%%%%%%%%%%%%%%%%%%%%%%%%%%%%%%%%%%%%%%%%%%%%%%%%%%%%%%%%%%%%%%%%%%%%%%$$
The curvature coefficient $\kappa_2^{S,\mu_{B}=0}$ as a function of both $G_{V}$ and $g_{V}$ are presented in Fig.~\ref{fig:kappa2_S}. The effect of $G_{V}/g_{V}$ on $\kappa_{2}^{S,\mu_{B}=0}$ is comparatively weaker than on $\kappa_{2}^{B,\mu_{S}=0}$ and $\kappa_{2}^{B,n_{s}=0}$. The dominant effect of both vector interactions $G_V, g_V$ on the light quark chiral transitions propagates through the respective number densities. For $\mu_{B}=0$, the light quark number densities get minor modifications due to $n_{s}\ne0$ leading to $\tilde{\mu}_{B}\ne0$ in Model-\RNum{1}; whereas they remain zero in Model-\RNum{2}. Hence, in Model-\RNum{2}, $g_{V}$ modifies only the strangeness chemical potential, which influences the chiral transition through the 't Hooft determinant term, resulting in a monotonically decreasing $\kappa_{2}^{S,\mu_{B}=0}$ as a function of $g_{V}$. On the other hand, Model-\RNum{1} is a little more subtle, as it induces a finite $\mu_{B}$, leading to a decrease in the crossover temperature. This reduces the effect of repulsive interaction (positive $G_{V}$) resulting in a slow variation of $\kappa_{2}^{S,\mu_{B}=0}$ as a function of $G_{V}$. On the other hand, for attractive interaction (negative $G_{V}$), the curvature increases at a faster rate. From the upper panel of Fig.~\ref{fig:kappa2_S}, the difference in slope in the positive and negative range of $G_{V}$ is clearly visible, where we have used identical $G_{V}/g_{V}$ ranges as used in Fig.~\ref{fig:kappa2_v}.

\begin{figure}[htb!]
\begin{center}
  \includegraphics[scale=0.56]{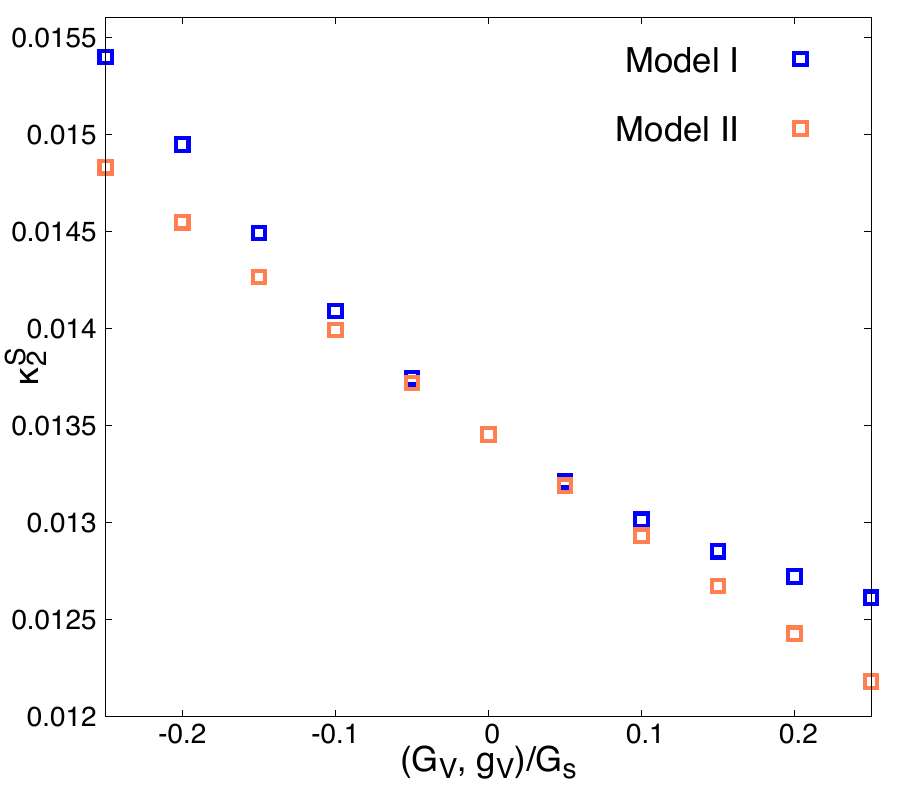}
  \caption{The curvature coefficients, $\kappa_{2}^{S,\mu_{B}=0}$ as a function of the vector interactions for both  models.}
  \label{fig:kappa2_S_com}
\end{center}  
\end{figure}

To compare Model-\RNum{1} and Model-\RNum{2}, we present the curvature $\kappa_{2}^{S,\mu_{B}=0}$ as a function of $G_{V}/g_{V}$ within a given range for both models in Fig.~\ref{fig:kappa2_S_com}. As already described, in Model-\RNum{2}, it has a monotonic decreasing behavior, whereas the effect of $G_{V}$ gets suppressed in the positive range and enhanced in the negative range for Model-\RNum{1}. It is to be noted that $\mu_{B}=0$ leads to $G_{V}\simeq g_{V}$, which is evident from Eq.~\eqref{eq:effective_mu}.

From Fig~\ref{fig:kappa2_S}, one might obtain an upper bound of $g_{V}$ in Model-\RNum{2}, but as presented in Table~\ref{tab:tablek2}, one can consider a different parameter set, mainly a different $K$ and shift these curves upward resulting in a different bound. Hence, we conclude that without a proper constraint on $K$, which can be obtained following the procedure presented in Ref.~\cite{Ali:2024nrz}, one cannot obtain a meaningful bound on the vector interactions from $\kappa_{2}^{S,\mu_{B}=0}$.

%%%%%%%%%%%%%%%%%%%%%%%%%%%%%%%%%%%%%%%%%%%%%%%%%%%%%%%%%%%%%%%%%%%%%%%%%%%

\clearpage

\bibliography{ref_common}

\end{document}